\documentclass[twocolumn,english,aps,prd]{revtex4-1}
\usepackage{CJK}
\usepackage[T1]{fontenc}
\usepackage[latin9]{inputenc}
\usepackage{geometry}
\usepackage{array}
\usepackage{textcomp}
\usepackage{multirow}
\usepackage{amsmath}
\usepackage{graphicx}
\usepackage{hhline}
\makeatletter

\providecommand{\tabularnewline}{\\}

\@ifundefined{date}{}{\date{}}

\usepackage{float}
\usepackage{textcomp}

\usepackage{babel}

\IfFileExists{lmodern.sty}{\usepackage{lmodern}}{}

\usepackage{babel}

\@ifundefined{showcaptionsetup}{}{%
 \PassOptionsToPackage{caption=false}{subfig}}
\usepackage{subfig}
\makeatother

\usepackage{babel}
\begin{document}
\begin{CJK*}{}{}
\title{$\boldsymbol{SU(3)_{C}\otimes SU(3)_{L}\otimes U(1)_{X}}$ models
in view of the $750$ GeV diphoton signal}

\author{R. Martinez}

\email{remartinezm@unal.edu.co}

\author{F. Ochoa}

\email{faochoap@unal.edu.co}

\author{C.F. Sierra}

\email{cfsierraf@unal.edu.co}

\affiliation{\textit{Departamento de F\'{i}sica, Universidad Nacional de Colombia,
Ciudad Universitaria, K. 45 No. 26-85, Bogot\'{a} D.C., Colombia} }
\begin{abstract}
We analyze the recent diphoton signal reported by ATLAS and CMS collaborations
in the context of the $SU(3)_{C}\otimes SU(3)_{L}\otimes U(1)_{X}$
anomaly free models, with a 750 GeV scalar candidate which can decay
into two photons. These models may explain the 750GeV signal by means
of one loop decays to $\gamma\gamma$ through charged vector and Higgs
bosons, as well as top-, bottom- and electron-like exotic particles
that arise naturally from the condition of anomaly cancellations of
the $SU(3)_{C}\otimes SU(3)_{L}\otimes U(1)_{X}$ models.
\end{abstract}
\maketitle
\end{CJK*}
\section{Introduction}

The recent excess in the 2015 ATLAS and CMS data with two photons
in the final state at invariant mass of about 750 GeV \cite{CMS750,ATLAS750}
has put under observation and testing a large number of models in
order to explain it (for a complete list of references see \cite{tools,Ellis}).

Particularly, we are interested in testing the models with gauge symmetry
$SU(3)_{C}\otimes SU(3)_{L}\otimes U(1)_{X},$ also called 331 models
\cite{Georgi-Pais,Pleitez,Frampton,Long,Pisano}. In these models,
after imposing some restrictions, as for example the cancellation
of anomalies, a free parameter $\beta$ remains and therefore it is
not possible to identify a unique version of a 331 model. The $\beta$
parameter determines the fermionic content of the model. For example,
for a given representation and $\beta=\sqrt{3}$ there appears exotic
quarks and leptons with electric charge $5/3$ and -2 times the proton
charge, respectively, while for $\beta=-1/\sqrt{3}$ there appears
new quarks of charge $2/3$ and extra neutrinos.

Recently, the diphoton excess analysis in the context of these models
has been addressed in references \cite{331diphoton}. Here we consider
the general case for $\beta=\pm\sqrt{3},$ and $\beta=\pm1/\sqrt{3}$
and two possible representations \cite{Georgi-Pais,Pleitez,Frampton}
taking into account possible interference effects between the new
vector and charged Higgs bosons that arise in the 331 models, which
can change considerably the production cross section. 

Since the 331 models require the three families in order to cancel
chiral anomalies\cite{anomalias}, these models arise as a possible
solution to the generation puzzle. They can also predict the charge
quantization for a three family model even when neutrino masses are
added \cite{Pires}. Also, in the framework of supersymmetric 331
models, the breaking chain GUT$\rightarrow331\rightarrow SM$ is allowed
and the model is protected from fast proton decay \cite{Martinez2}.
In addition, recent versions of the model have addressed the mass
hierarchy problem both in the quark and lepton sectors \cite{martinez2006,Dias,Dong,Martinez-Carcamo,Long-1,Valle}
as well as the dark matter problem \cite{Mizukoshi,Dias2010,Alvares2012,Cogollo2014,Queiroz-Martinez}.

However, there are some features that neither SM \cite{SM} nor the
331 extensions have been able to explain at a cosmological level,
such as the formation of large scale structures in the universe \cite{Volkas},
the origin of the galactic halo \cite{Mohapatra}, and the observations
of gamma ray bursts \cite{Wong}. On the other hand, the model is
purely left-handed, so that it cannot account for the parity breaking.
Another point of interest to study in these models is the CP violation,
particularly the strong CP violation which might allow us to understand
the values of the electric dipole moment of the neutron and electron
\cite{Pal,Dias-Pleitez-pires}.

This paper is organized as follows. In section 2, we review the main
features of the 331 models, their spontaneous symmetry breaking (SSB)
scheme, their Higgs potentials as well as the Yukawa Lagrangians with
the relevant particle content resulting from the $\beta$ parameter
choice. Then, in section 3, we study the diphoton decay in the framework
of the 331 models for $\beta=\pm\sqrt{3},$ and $\beta=\pm1/\sqrt{3}$,
finding restrictions for each case consistent with the reported cross
section of the 750 GeV signal.

\section{Description of the model}

Although cancellation of anomalies leads to some conditions \cite{fourteen},
such criterion alone still permits an infinite number of 331 models.
In these models, the electric charge is defined in general as a linear
combination of the diagonal generators of the group 
\begin{equation}
Q=T_{3}+\beta T_{8}+XI,\label{charge}
\end{equation}

\noindent with $T_{3}=\frac{1}{2}diag(1,-1,0)$ and $T_{8}=\frac{1}{2\sqrt{3}}diag(1,1,-2),$
$I=diag\left(1,1,1\right)$ is the identity matrix and $X$ is the
quantum number associated to the $U(1)_{X}$ group. The study of $\beta$
is interesting because it determines the fermion assignment, and more
specifically, the electric charges of the extra particle sector. We
consider the most popular models for $\beta=\pm\sqrt{3},$ and $\beta=\pm1/\sqrt{3}\ $\cite{Georgi-Pais,Pleitez,Frampton,Martinez-Diaz-Ochoa}.
Here, we assume the following symmetry breaking pattern

\begin{widetext}

\begin{eqnarray*}
SU\left(3\right)_{C}\otimes SU\left(3\right)_{L}\otimes U\left(1\right)_{X} & \rightarrow & SU\left(3\right)_{C}\otimes SU\left(2\right)_{L}\otimes U\left(1\right)_{Y}\rightarrow SU\left(3\right)_{C}\otimes U\left(1\right)_{Q},\\
331 & \rightarrow & 321\rightarrow31.
\end{eqnarray*}

\end{widetext}

Although the spontaneous symmetry breaking of the group is possible
with less than three scalar triplets, this option does not allow a
Peccei-Quinn symmetry in order to face the strong-CP problem \cite{Peccei-Dong}.
So, we use one scalar triplet for the first symmetry breaking and
two scalar triplets for the second to give masses to the up and down
sectors of the SM (see Table \ref{tab:Scalar-spectrum}). The triplet
field $\chi$ only introduce a VEV on the third component for the first
transition and  induces the masses of the exotic fermionic components.
In the second transition pairs of solutions are obtained according
to the value of $\beta$. A detailed analysis of such solutions shows
that two multiplets are necessary in order to give masses to
the quarks of type up and down simultaneously \cite{Martinez-Diaz-Ochoa}.
Therefore, we introduce two triplets$\;\rho$ and
$\eta$ in the second transition. In some cases a scalar sextet is
introduced to give masses to the neutrinos \cite{Pleitez}.

\begin{widetext}

\begin{table}[!h]
\begin{centering}
{\small{}}%
\begin{tabular}{ccc}
\hhline{===} 
 Spectrum & $\mathit{SU(3)_{L}\otimes U(1)_{X}}$ & $\mathit{Q}$\tabularnewline
\hline 
$\chi=\begin{pmatrix}\chi_{1}^{\mathit{\pm Q_{1}}}\\
\chi_{2}^{\mathit{\pm Q_{2}}}\\
\frac{1}{\sqrt{2}}(\upsilon_{\chi}+\xi_{\chi}\pm i\zeta_{\chi})
\end{pmatrix}$ & $\left(\mathbf{3},\frac{\beta}{\sqrt{3}}\right)$ & $\begin{pmatrix}\pm\left(\frac{1}{2}+\frac{\sqrt{3}\beta}{2}\right)\\
\pm\left(-\frac{1}{2}+\frac{\sqrt{3}\beta}{2}\right)\\
0
\end{pmatrix}$\tabularnewline
$\rho=\begin{pmatrix}\rho_{1}^{\pm}\\
\frac{1}{\sqrt{2}}(\upsilon_{\rho}+\xi_{\rho}\pm i\zeta_{\rho})\\
\rho_{3}^{\mathit{\mp Q_{2}}}
\end{pmatrix}$ & $\left(\mathbf{3},\frac{1}{2}-\frac{\beta}{2\sqrt{3}}\right)$ & $\begin{pmatrix}\pm1\\
0\\
\mp\left(-\frac{1}{2}+\frac{\sqrt{3}\beta}{2}\right)
\end{pmatrix}$\tabularnewline
$\eta=\begin{pmatrix}\frac{1}{\sqrt{2}}(\upsilon_{\eta}+\xi_{\eta}\pm i\zeta_{\eta})\\
\eta_{2}^{\mp}\\
\eta_{3}^{\mp\mathit{Q_{1}}}
\end{pmatrix}$ & $\left(\mathbf{3},-\frac{1}{2}-\frac{\beta}{2\sqrt{3}}\right)$ & $\begin{pmatrix}0\\
\mp1\\
\mp\left(\frac{1}{2}+\frac{\sqrt{3}\beta}{2}\right)
\end{pmatrix}$\tabularnewline
\hhline{===} 
\end{tabular}
\par\end{centering}{\small \par}

\caption{{\small{}Scalar spectrum for the SSB $331\rightarrow321\rightarrow31$.
Here} $Q_{1}=\frac{1}{2}+\frac{\sqrt{3}\beta}{2}$ and $Q_{2}=-\frac{1}{2}+\frac{\sqrt{3}\beta}{2}$.{\small{}\label{tab:Scalar-spectrum}}}

{\small{}.}{\small \par}
\end{table}

\begin{table}[!h]
\begin{centering}
{\small{}}%
\begin{tabular}{ccc}
\hhline{===} 
$\beta$ & $V_{\xi_{\chi}}$ & $\mathcal{L}_{\xi_{\chi}}^{HVV}$\tabularnewline
\hline 
$\frac{1}{\sqrt{3}}$ & $\lambda_{7}\upsilon_{\chi}\xi_{\chi}\eta_{3}^{+}\eta_{3}^{-}$ & $\frac{g_{L}m_{K}}{\sqrt{2}}\xi_{\chi}K^{\mu\pm}K_{\mu}^{\mp}$\tabularnewline
$-\frac{1}{\sqrt{3}}$ & $\lambda_{8}\upsilon_{\chi}\xi_{\chi}\rho_{3}^{+}\rho_{3}^{-}$ & $\frac{g_{L}m_{K}}{\sqrt{2}}\xi_{\chi}K^{\mu\pm}K_{\mu}^{\mp}$\tabularnewline
$\sqrt{3}$ & $\qquad\upsilon_{\chi}\xi_{\chi}\left(\lambda_{7}\eta_{3}^{++}\eta_{3}^{--}+\lambda_{8}\rho_{3}^{+}\rho_{3}^{-}\right)\qquad$ & $\frac{g_{L}\xi_{\chi}}{\sqrt{2}}\left(m_{K^{++}}K^{\mu\pm\pm}K_{\mu}^{\mp\mp}+m_{K}K^{\mu\pm}K_{\mu}^{\mp}\right)$\tabularnewline
$-\sqrt{3}$ & $\upsilon_{\chi}\xi_{\chi}\left(\lambda_{7}\eta_{3}^{+}\eta_{3}^{-}+\lambda_{8}\rho_{3}^{++}\rho_{3}^{--}\right)$ & $\qquad\frac{g_{L}\xi_{\chi}}{\sqrt{2}}\left(m_{K^{--}}K^{\mu\mp}K_{\mu}^{\pm}+m_{K}K^{\mu\mp}K_{\mu}^{\pm}\right)\qquad$\tabularnewline
\hhline{===} 
\end{tabular}
\par\end{centering}{\small \par}

\caption{Relevant bosonic trilinear couplings with $\xi_{\chi}$.{\small{}\label{tab:trilinear}}}
\end{table}

\subsection{Bosonic sector}

The most general and renormalizable form of the Higgs potential, taking
into account all the possible linear combinations among the three
triplets forming quadratic, cubic, and quartic products invariant
under $SU(3)_{L}\otimes U(1)_{X}$ is given by \cite{Martinez-Diaz-Ochoa}:
\begin{enumerate}
\item For $\beta=\frac{1}{\sqrt{3}}$
\begin{align}
V & =\mu_{1}^{2}\chi^{i}\chi_{i}+\mu_{2}^{2}\rho^{i}\rho_{i}+\mu_{3}^{2}\eta^{i}\eta_{i}+\mu_{4}^{2}\left(\chi^{i}\rho_{i}+h.c\right)+f\left(\chi_{i}\rho_{j}\eta_{k}\varepsilon^{ijk}+h.c\right)\notag\\
 & +\lambda_{1}(\chi^{i}\chi_{i})^{2}+\lambda_{2}(\rho^{i}\rho_{i})^{2}+\lambda_{3}(\eta^{i}\eta_{i})^{2}+\lambda_{4}\chi^{i}\chi_{i}\rho^{j}\rho_{j}+\lambda_{5}\chi^{i}\chi_{i}\eta^{j}\eta_{j}\notag\\
 & +\lambda_{6}\rho^{i}\rho_{i}\eta^{j}\eta_{j}+\lambda_{7}\chi^{i}\eta_{i}\eta^{j}\chi_{j}+\lambda_{8}\chi^{i}\rho_{i}\rho^{j}\chi_{j}+\lambda_{9}\eta^{i}\rho_{i}\rho^{j}\eta_{j}\notag\\
 & +\lambda_{10}\chi^{i}\chi_{i}\left(\chi^{j}\rho_{j}+h.c.\right)+\lambda_{11}\rho^{i}\rho_{i}\left(\rho^{j}\chi_{j}+h.c.\right)+\lambda_{12}\eta^{i}\eta_{i}\left(\chi^{j}\rho_{j}+h.c.\right)\notag\\
 & +\lambda_{13}\left(\chi^{i}\rho_{i}\chi^{j}\rho_{j}+h.c.\right)+\lambda_{14}\left(\eta^{i}\chi_{i}\rho^{j}\eta_{j}+h.c.\right).\label{21a}
\end{align}

\item For $\beta=-\frac{1}{\sqrt{3}}$
\begin{align}
V & =\mu_{1}^{2}\chi^{i}\chi_{i}+\mu_{2}^{2}\rho^{i}\rho_{i}+\mu_{3}^{2}\eta^{i}\eta_{i}+\mu_{4}^{2}\left(\chi^{i}\eta_{i}+h.c\right)+f\left(\chi_{i}\rho_{j}\eta_{k}\varepsilon^{ijk}+h.c\right)\notag\\
 & +\lambda_{1}(\chi^{i}\chi_{i})^{2}+\lambda_{2}(\rho^{i}\rho_{i})^{2}+\lambda_{3}(\eta^{i}\eta_{i})^{2}+\lambda_{4}\chi^{i}\chi_{i}\rho^{j}\rho_{j}+\lambda_{5}\chi^{i}\chi_{i}\eta^{j}\eta_{j}\notag\\
 & +\lambda_{6}\rho^{i}\rho_{i}\eta^{j}\eta_{j}+\lambda_{7}\chi^{i}\eta_{i}\eta^{j}\chi_{j}+\lambda_{8}\chi^{i}\rho_{i}\rho^{j}\chi_{j}+\lambda_{9}\eta^{i}\rho_{i}\rho^{j}\eta_{j}\notag\\
 & +\lambda_{10}\chi^{i}\chi_{i}\left(\chi^{j}\eta_{j}+h.c.\right)+\lambda_{11}\eta^{i}\eta_{i}\left(\eta^{j}\chi_{j}+h.c.\right)+\lambda_{12}\rho^{i}\rho_{i}\left(\chi^{j}\eta_{j}+h.c.\right)\notag\\
 & +\lambda_{13}\left(\chi^{i}\eta_{i}\chi^{j}\eta_{j}+h.c.\right)+\lambda_{14}\left(\rho^{i}\chi_{i}\eta^{j}\rho_{j}+h.c.\right).\label{potbeta2}
\end{align}

\item For $\beta=\sqrt{3}$
\begin{align}
V & =\mu_{1}^{2}\chi^{i}\chi_{i}+\mu_{2}^{2}\rho^{i}\rho_{i}+\mu_{3}^{2}\eta^{i}\eta_{i}+f\left(\chi_{i}\rho_{j}\eta_{k}\varepsilon^{ijk}+h.c\right)+\lambda_{1}(\chi^{i}\chi_{i})^{2}+\lambda_{2}(\rho^{i}\rho_{i})^{2}\notag\\
 & +\lambda_{3}(\eta^{i}\eta_{i})^{2}+\lambda_{4}\chi^{i}\chi_{i}\rho^{j}\rho_{j}+\lambda_{5}\chi^{i}\chi_{i}\eta^{j}\eta_{j}+\lambda_{6}\rho^{i}\rho_{i}\eta^{j}\eta_{j}+\lambda_{7}\chi^{i}\eta_{i}\eta^{j}\chi_{j}\notag\\
 & +\lambda_{8}\chi^{i}\rho_{i}\rho^{j}\chi_{j}+\lambda_{9}\eta^{i}\rho_{i}\rho^{j}\eta_{j}+\lambda_{10}\left(\rho^{i}\chi_{i}\rho^{j}\eta_{j}+h.c.\right).\label{potbeta3}
\end{align}

\item For $\beta=-\sqrt{3}$
\begin{align}
V & =\mu_{1}^{2}\chi^{i}\chi_{i}+\mu_{2}^{2}\rho^{i}\rho_{i}+\mu_{3}^{2}\eta^{i}\eta_{i}+f\left(\chi_{i}\rho_{j}\eta_{k}\varepsilon^{ijk}+h.c\right)+\lambda_{1}(\chi^{i}\chi_{i})^{2}+\lambda_{2}(\rho^{i}\rho_{i})^{2}\notag\\
 & +\lambda_{3}(\eta^{i}\eta_{i})^{2}+\lambda_{4}\chi^{i}\chi_{i}\rho^{j}\rho_{j}+\lambda_{5}\chi^{i}\chi_{i}\eta^{j}\eta_{j}+\lambda_{6}\rho^{i}\rho_{i}\eta^{j}\eta_{j}+\lambda_{7}\chi^{i}\eta_{i}\eta^{j}\chi_{j}\notag\\
 & +\lambda_{8}\chi^{i}\rho_{i}\rho^{j}\chi_{j}+\lambda_{9}\eta^{i}\rho_{i}\rho^{j}\eta_{j}+\lambda_{10}\left(\eta^{i}\chi_{i}\eta^{j}\rho_{j}+h.c.\right).\label{potbeta4}
\end{align}

\end{enumerate}
\end{widetext}

The rotation matrices to mass eigenvectors will have the standard
form

\begin{equation}
\left(\begin{array}{c}
\eta_{2}^{\pm}\\
\rho_{1}^{\pm}
\end{array}\right)=\left(\begin{array}{cc}
C_{\beta} & S_{\beta}\\
-S_{\beta} & C_{\beta}
\end{array}\right)\left(\begin{array}{c}
G^{\pm}\\
H^{\pm}
\end{array}\right),\label{eq:eigencharg}
\end{equation}

\begin{equation}
\left(\begin{array}{c}
\xi_{\rho}\\
\xi_{\eta}\\
\xi_{\chi}
\end{array}\right)=\left(\begin{array}{ccc}
C_{\alpha} & S_{\alpha} & 0\\
-S_{\alpha} & C_{\alpha} & 0\\
0 & 0 & 1
\end{array}\right)\left(\begin{array}{c}
h\\
H\\
H_{3}
\end{array}\right).\label{eq:eigenneu}
\end{equation}

We take the real component $\xi_{\chi}$ from the field $\chi$ as
our 750 GeV signal candidate, corresponding to one of the residual
physical particles after the $SU\left(3\right)_{L}\otimes U\left(1\right)_{X}$
symmetry breaking, while the imaginary component $\zeta_{\chi}$ corresponds
to the would-be Goldstone boson that become into the longitudinal
component of a $Z'$ gauge boson. So, after rotation to mass eigenvectors
according to Eqs.(\ref{eq:eigencharg}-\ref{eq:eigenneu}), we obtain
all the interactions of $\xi_{\chi}$ with the scalar matter in the
framework of an effective Two Higgs Doublet Model (2HDM) in the low
energy limit, where both electroweak triplets $\rho$ and $\eta$
are decomposed into two hypercharge-one $SU(2)_{L}$ doublets plus
charged and neutral singlets. In particular, the masses of the extra
neutral, pseudoscalar and charged Higgs bosons $H$, $A$ and $H^{\pm}$,
respectively, are nearly degenerated and at the TeV scale, as shown
in \cite{Martinez-Diaz-Ochoa}. So, the decay of $\xi_{\chi}$ into
these Higgs bosons are kinematically forbidden. Explicitly, the couplings
with the resulting charged Higgs boson $H^{\pm}$ of the 2HDM are
given by

\begin{widetext}

\begin{align}
V_{\xi_{\chi}}^{\mathrm{2HDM}} & =\upsilon_{\chi}\xi_{\chi}\left[\left(\lambda_{4}C_{\beta}^{2}+\lambda_{5}S_{\beta}^{2}\right)H^{+}H^{-}+\frac{1}{2}\left(\lambda_{4}C_{\alpha}^{2}+\lambda_{5}S_{\alpha}^{2}\right)h^{2}\right.\nonumber \\
 & \left.+\frac{1}{2}\left(\lambda_{4}S_{\alpha}^{2}+\lambda_{5}C_{\alpha}^{2}\right)H^{2}+\left(\lambda_{4}-\lambda_{5}\right)C_{\alpha}S_{\alpha}hH\right].\label{eq:2hdm-1}
\end{align}

For the 331 models, in the limit $\upsilon_{\chi}^{2}>>\upsilon_{\rho}^{2},\upsilon_{\eta}^{2}$
we obtain the relation $\alpha\approx\beta\pm\frac{\pi}{2}$ \cite{Martinez-Diaz-Ochoa}
allowing us to simplify the previous expression to

\begin{align}
V_{\xi_{\chi}}^{\mathrm{2HDM}} & =\upsilon_{\chi}\xi_{\chi}\left[\lambda\left(H^{+}H^{-}+\frac{h^{2}+H^{2}}{2}\right)+\left(\lambda_{4}-\lambda_{5}\right)C_{\alpha}S_{\alpha}hH\right]
\end{align}

where we have defined $\lambda\equiv\lambda_{4}S_{\alpha}^{2}+\lambda_{5}C_{\alpha}^{2}$.
Also, as a particular case, if we had set $\lambda_{4}=\lambda_{5}$
in Eq.(\ref{eq:2hdm-1}), we would have obtained the same coupling
for the decay $\xi_{\chi}\rightarrow hh$ and $\xi_{\chi}\rightarrow H^{+}H^{-}$
independently on the mixing angles. In this way, since the decay $\xi_{\chi}\rightarrow hh$
is strongly constrained by ATLAS and CMS at 95\%CL \cite{Ellis},
the coupling between $\xi_{\chi}$ and $H^{\pm}$ is also suppressed,
thus the charged Higgs boson $H^{\pm}$ will not contribute to the
diphoton decay.

On the other hand, the relevant trilinear couplings with the extra
vector bosons $K^{\pm Q_{1}}$ and $K^{\pm Q_{2}}$ are given by \cite{Martinez-Diaz-Ochoa}

\begin{equation}
\mathcal{L}_{\xi_{\chi}}^{HVV}=\frac{g_{L}\xi_{\chi}}{\sqrt{2}}\left(m_{K^{Q_{1}}}K^{\mu\pm Q_{1}}K_{\mu}^{\mp Q_{1}}+m_{K^{Q_{2}}}K^{\mu\pm Q_{2}}K_{\mu}^{\mp Q_{2}}\right).
\end{equation}

\noindent where $g_{L}$ is the $SU(2)_{L}$ coupling constant. Taking
into account all the above conditions and after the $SU\left(3\right)_{L}\otimes U\left(1\right)_{X}$
symmetry breaking by $\upsilon_{\chi}$, we obtain the relevant trilinear
bosonic couplings with the third components of the triplets which
correspond to singlets fields under the SM. According to the $\beta$
value, these singlet fields can be charged or doubly charged, contributing
to the diphoton decay according to the Feynman rules in Table \ref{tab:trilinear}.
In the loops we will refer to these fields as $h^{\pm}$ and $h^{\pm\pm}$,
respectively.

\subsection{Fermionic sector}

The fermions exhibit the following general structure of transformations
under the chiral group $SU(3)_{L}\otimes U(1)_{X}$

\begin{eqnarray}
\psi_{L} & = & \left\{ \begin{array}{c}
q_{L}:\left(\mathbf{3,}X_{q}^{L}\right)=\left(\mathbf{2,}X_{q}^{L}\right)\oplus\left(\mathbf{1,}X_{q}^{L}\right),\\
\ell_{L}:\left(\mathbf{3,}X_{\ell}^{L}\right)=\left(\mathbf{2,}X_{\ell}^{L}\right)\oplus\left(\mathbf{1,}X_{\ell}^{L}\right),
\end{array}\right.\notag\\
\psi_{L}^{\ast} & = & \left\{ \begin{array}{c}
q_{L}^{\ast}:\left(\mathbf{3}^{\ast}\mathbf{,-}X_{q}^{L}\right)=\left(\mathbf{2}^{\ast}\mathbf{,-}X_{q}^{L}\right)\oplus\left(\mathbf{1,-}X_{q}^{L}\right)\mathbf{,}\\
\ell_{L}^{\ast}:\left(\mathbf{3}^{\ast}\mathbf{,-}X_{\ell}^{L}\right)=\left(\mathbf{2}^{\ast}\mathbf{,-}X_{\ell}^{L}\right)\oplus\left(\mathbf{1,-}X_{\ell}^{L}\right),
\end{array}\right.\notag\\
\psi_{R} & = & \left\{ \begin{array}{c}
q_{R}:\left(\mathbf{1,}X_{q}^{R}\right),\\
\ell_{R}:\left(\mathbf{1,}X_{\ell}^{R}\right),
\end{array}\right.\label{1}
\end{eqnarray}
where the quarks $q$ can be either color triplets (\textbf{3}) or
antitriplets ($\boldsymbol{3^{*}}$) according to the representation
choice and the leptons $\ell$ are color singlets (\textbf{1}). The
second equality corresponds to the branching rules $SU(2)_{L}\subset SU(3)_{L}$.
The possibilities \textbf{$\mathbf{3}$ }and $\boldsymbol{3^{*}}$
are included in both the color and flavor sector since the same number
of fermion triplets and antitriplets must be present in order to cancel
anomalies \cite{doff} and the quantum number $X$, associated with
the parameter $\beta$, will take specific values according to the
representations of $SU(3)_{L}$ and the anomalies cancellation. In
this way, there are two possible representations for the 331 models
that are mutually conjugated, that we call models A and A{*}. The
difference between the two models is to change $\beta$ in model A
by $-\beta$ in model A{*} (Tables \ref{tab:fermioncontent}-\ref{tab:fermionsloop}).
Henceforth we will use the model A in order to evaluate the production
cross section.

\begin{table}[!tbh]
\begin{centering}
\begin{tabular}{cccc}
\hhline{====} 
Model & Spectrum & $SU\left(3\right)_{L}\otimes U\left(1\right)_{X}$ & $\mathit{Q}$\tabularnewline
\hline 
\multirow{11}{*}{A} & $\ensuremath{\begin{tabular}{c}
 \ensuremath{q_{L}^{3}=\left(\begin{array}{c}
U^{3}\\
D^{3}\\
T^{3}(J^{3})
\end{array}\right)_{L}} \\
 
\end{tabular}}$ & $\left(\ensuremath{\mathbf{3},\frac{1}{6}-\frac{\beta}{2\sqrt{3}}}\right)$ & $\ensuremath{\left(\begin{array}{c}
\frac{2}{3}\\
-\frac{1}{3}\\
\frac{1}{6}-\frac{\sqrt{3}\beta}{2}
\end{array}\right)}$\tabularnewline
 &  &  & \tabularnewline
 & $\ensuremath{U_{R}^{3}}$,$\ensuremath{D_{R}^{3}}$,$\ensuremath{T_{R}^{3}}$,$(J_{R}^{3})$ & $\qquad\left(\ensuremath{\mathbf{1},\frac{2}{3}}\right)$,$\left(\ensuremath{\mathbf{1},-\frac{1}{3}}\right)$,$\left(\ensuremath{\mathbf{1},\frac{1}{6}-\frac{\sqrt{3}\beta}{2}}\right)\qquad$ & $\ensuremath{\frac{2}{3}}$,$\ensuremath{-\frac{1}{3}}$,$\frac{1}{6}-\frac{\sqrt{3}\beta}{2}$\tabularnewline
 &  &  & \tabularnewline
 & $\begin{tabular}{c}
 \ensuremath{q_{L}^{1,2}=\left(\begin{array}{c}
D^{1,2}\\
-U^{1,2}\\
J^{1,2}(T^{1,2})
\end{array}\right)_{L}}\\
 
\end{tabular}$ & $\left(\ensuremath{\boldsymbol{3^{*}},\frac{1}{6}+\frac{\beta}{2\sqrt{3}}}\right)$ & $\ensuremath{\left(\begin{array}{c}
-\frac{1}{3}\\
\frac{2}{3}\\
\frac{1}{6}-\frac{\sqrt{3}\beta}{2}
\end{array}\right)}$\tabularnewline
 &  &  & \tabularnewline
 & $\ensuremath{D_{R}^{1,2}}$,$\ensuremath{U_{R}^{1,2}}$,$J_{R}^{1,2}(T^{1,2})$ & $\qquad\left(\ensuremath{\mathbf{1},-\frac{1}{3}}\right)$,$\left(\ensuremath{\mathbf{1},\frac{2}{3}}\right)$,$\left(\ensuremath{\mathbf{1},\frac{1}{6}+\frac{\sqrt{3}\beta}{2}}\right)\qquad$ & $-\ensuremath{\frac{1}{3}}$,$\ensuremath{\frac{2}{3}}$,$\frac{1}{6}+\frac{\sqrt{3}\beta}{2}$\tabularnewline
 &  &  & \tabularnewline
 & $\ensuremath{\ell_{L}^{(n)}=\left(\begin{array}{c}
\nu^{n}\\
e^{n}\\
E^{n}
\end{array}\right)_{L}}$ & $\left(\ensuremath{\ensuremath{\mathbf{3},-\frac{1}{2}-\frac{\beta}{2\sqrt{3}}}}\right)$ & $\ensuremath{\left(\begin{array}{c}
0\\
-1\\
-\frac{1}{2}-\frac{\sqrt{3}\beta}{2}
\end{array}\right)}$\tabularnewline
 &  &  & \tabularnewline
 & $\ensuremath{\nu_{R}^{n}}$,$\mathit{e_{R}^{n}}$,$\mathit{E_{R}^{n}}$ & $\qquad\left(\mathbf{1},0\right)$,$\left(\mathbf{1},-1\right)$,$\left(\mathbf{1},-\frac{1}{2}-\frac{\sqrt{3}\beta}{2}\right)\qquad$ & $0$,$-1$,$-\frac{1}{2}-\frac{\sqrt{3}\beta}{2}$\tabularnewline
\hline 
\multirow{12}{*}{A{*}} &  &  & \tabularnewline
 & $\ensuremath{\begin{tabular}{c}
 \ensuremath{q_{L}^{3}=\left(\begin{array}{c}
D^{3}\\
-U^{3}\\
J^{3}(T^{3})
\end{array}\right)_{L}} \\
 
\end{tabular}}$ & $\left(\ensuremath{\boldsymbol{3^{*}},\frac{1}{6}+\frac{\beta}{2\sqrt{3}}}\right)$ & $\ensuremath{\left(\begin{array}{c}
-\frac{1}{3}\\
\frac{2}{3}\\
\frac{1}{6}+\frac{\sqrt{3}\beta}{2}
\end{array}\right)}$\tabularnewline
 &  &  & \tabularnewline
 & $\ensuremath{D_{R}^{3}}$,$\ensuremath{U_{R}^{3}}$,$J_{R}^{3}$$(\ensuremath{T_{R}^{3})}$ & $\qquad\left(\ensuremath{\mathbf{1},-\frac{1}{3}}\right)$,$\left(\ensuremath{\mathbf{1},\frac{2}{3}}\right)$,$\left(\ensuremath{\mathbf{1},\frac{1}{6}+\frac{\sqrt{3}\beta}{2}}\right)\qquad$ & $-\ensuremath{\frac{1}{3}}$,$\ensuremath{\frac{2}{3}}$,$\frac{1}{6}+\frac{\sqrt{3}\beta}{2}$\tabularnewline
 &  &  & \tabularnewline
 & $\begin{tabular}{c}
 \ensuremath{q_{L}^{1,2}=\left(\begin{array}{c}
U^{1,2}\\
D^{1,2}\\
T^{1,2}(J^{1,2})
\end{array}\right)_{L}} \\
 
\end{tabular}$ & $\left(\ensuremath{\mathbf{3},\frac{1}{6}-\frac{\beta}{2\sqrt{3}}}\right)$ & $\ensuremath{\left(\begin{array}{c}
\frac{2}{3}\\
-\frac{1}{3}\\
\frac{1}{6}+\frac{\sqrt{3}\beta}{2}
\end{array}\right)}$\tabularnewline
 &  &  & \tabularnewline
 & $\ensuremath{U_{R}^{1,2}}$,$\ensuremath{D_{R}^{1,2}}$,$T^{1,2}(J_{R}^{1,2})$ & $\qquad\left(\ensuremath{\mathbf{1},\frac{2}{3}}\right)$,$\left(\ensuremath{\mathbf{1},-\frac{1}{3}}\right)$,$\left(\ensuremath{\mathbf{1},\frac{1}{6}-\frac{\sqrt{3}\beta}{2}}\right)\qquad$ & $\ensuremath{\frac{2}{3}}$,$\ensuremath{-\frac{1}{3}}$,$\frac{1}{6}-\frac{\sqrt{3}\beta}{2}$\tabularnewline
 &  &  & \tabularnewline
 & $\ensuremath{\ell_{L}^{(n)}=\left(\begin{array}{c}
e^{n}\\
-\nu^{n}\\
E^{n}
\end{array}\right)_{L}}$ & $\left(\ensuremath{\ensuremath{\boldsymbol{3^{*}}-\frac{1}{2}+\frac{\beta}{2\sqrt{3}}}}\right)$ & $\ensuremath{\left(\begin{array}{c}
-1\\
0\\
-\frac{1}{2}+\frac{\sqrt{3}\beta}{2}
\end{array}\right)}$\tabularnewline
 &  &  & \tabularnewline
 & $\mathit{e_{R}^{n}}$,$\ensuremath{\nu_{R}^{n}}$,$\mathit{E_{R}^{n}}$ & $\qquad\left(\mathbf{1},-1\right)$,$\left(\mathbf{1},0\right)$,$\left(\mathbf{1},-\frac{1}{2}+\frac{\sqrt{3}\beta}{2}\right)\qquad$ & $-1$,$0$,$-\frac{1}{2}+\frac{\sqrt{3}\beta}{2}$\tabularnewline
\hhline{====} 
\end{tabular}
\par\end{centering}

\caption{Particle content for the fermionic sector with\textit{ $n=1,2,3$.
}The choice of $T$ or $J$ in the quark triplets depends on the value
of $\beta$.}
\label{tab:fermioncontent} 
\end{table}

\begin{table}[!h]
\begin{centering}
{\small{}}%
\begin{tabular}{ccccc}
\hhline{=====} 
 & \multicolumn{2}{c}{$\textrm{Fermions in the loop}$} & \multicolumn{2}{c}{$Q$}\tabularnewline
\cline{2-5} 
$\beta$ & Model A & Model A{*} & Model A & Model A{*}\tabularnewline
\hline 
$\frac{1}{\sqrt{3}}$ &  $\qquad T^{m}$, $J$, $E^{-}\qquad$  & $T$, $J^{m}$  & $\frac{2}{3}$, $-\frac{1}{3}$,$-1$ & $\frac{2}{3}$, $-\frac{1}{3}$\tabularnewline
$-\frac{1}{\sqrt{3}}$ & $T$, $J^{m}$  & $\qquad T^{m}$, $J$, $E^{-}\qquad$ & $\frac{2}{3}$, $-\frac{1}{3}$ & $\frac{2}{3}$, $-\frac{1}{3}$,$-1$\tabularnewline
$\sqrt{3}$ & $T^{m}$, $J$, $E^{--}$  & $T$, $J^{m}$, $E^{+}$  & $\qquad\frac{5}{3}$, $-\frac{4}{3}$, $-2\qquad$ & $\frac{2}{3}$, $-\frac{4}{3}$, $+1$\tabularnewline
$-\sqrt{3}$ & $T$, $J^{m}$, $E^{+}$  & $T^{m}$, $J$, $E^{--}$  & $\frac{2}{3}$, $-\frac{4}{3}$, $+1$ & $\qquad\frac{5}{3}$, $-\frac{4}{3}$, $-2\qquad$\tabularnewline
\hhline{=====}  
\end{tabular}
\par\end{centering}{\small \par}

\caption{Fermions in the loop for every choice of $\beta$.{\small{} Here $m=1,\,2$.\label{tab:fermionsloop}}}
\end{table}

Regardless the fermionic content of the model, the $\beta$
parameter and the representation choice,
the most general, renormalizable, and $SU\left(3\right)_{C}\otimes SU\left(3\right)_{L}\otimes U\left(1\right)_{X}$
invariant Yukawa Lagrangian for quarks is given by

\begin{eqnarray}
-\mathcal{L}_{Y}^{q} & = & \sum_{m=1}^{2}\overline{q_{L}^{3}}\left(h_{U_{R}^{3}}^{3\eta}U_{R}^{3}\eta+h_{D_{R}^{3}}^{3\rho}D_{R}^{3}\rho+h_{J_{R}^{3}}^{3\chi}J_{R}^{3}\chi+h_{U_{R}^{m}}^{3\eta}U_{R}^{3}\eta+h_{D_{R}^{m}}^{3\rho}D_{R}^{m}\rho\right)\nonumber \\
 & + & \sum_{m,m\text{\textasciiacute}=1}^{2}\overline{q_{L}^{m}}\left(h_{D_{R}^{3}}^{m\eta}D_{R}^{3}\eta+h_{U_{R}^{3}}^{m\rho}U_{R}^{3}\rho+h_{D_{R}^{m\text{\textasciiacute}}}^{m\eta}D_{R}^{m\text{\textasciiacute}}\eta+h_{U_{R}^{m\text{\textasciiacute}}}^{m\rho}U_{R}^{m\text{\textasciiacute}}\rho+h_{J_{R}^{m\text{\textasciiacute}}}^{m\chi}J_{R}^{m\text{\textasciiacute}}\chi\right)\nonumber \\
 & + & \mathcal{L}_{\pm1/\sqrt{3}}^{q}+h.c.\label{eq:yuakwaquarks}
\end{eqnarray}

where $\mathcal{L}_{\pm1/\sqrt{3}}^{q}$ contains mixing terms between
the SM light quarks and the exotic quarks $T$ and $J$ given by

\begin{align}
\mathcal{L}_{1/\sqrt{3}}^{q}= & \sum_{m=1}^{2}\overline{q_{L}^{3}}\left(h_{D_{R}^{3}}^{3\chi}D_{R}^{3}\chi+h_{J_{R}^{3}}^{3\rho}J_{R}^{3}\rho+h_{D_{R}^{m}}^{3\chi}D_{R}^{m}\chi+h_{J_{R}^{m}}^{3\eta}J_{R}^{m}\eta\right)\nonumber \\
+ & \sum_{m,m\text{\textasciiacute}=1}^{2}\overline{q_{L}^{m}}\left(h_{U_{R}^{3}}^{m\chi}U_{R}^{3}\chi+h_{J_{R}^{3}}^{m\eta}J_{R}^{3}\eta+h_{U_{R}^{m\text{\textasciiacute}}}^{m\chi}U_{R}^{m\text{\textasciiacute}}\chi+h_{J_{R}^{m\text{\textasciiacute}}}^{m\rho}J_{R}^{m\text{\textasciiacute}}\rho\right)+h.c.
\end{align}

\begin{align}
\mathcal{L}_{-1/\sqrt{3}}^{q}= & \sum_{m=1}^{2}\overline{q_{L}^{3}}\left(h_{U_{R}^{3}}^{3\chi}U_{R}^{3}\chi+h_{J_{R}^{3}}^{3\eta}J_{R}^{3}\eta+h_{U_{R}^{m}}^{3\chi}U_{R}^{m}\chi+h_{J_{R}^{m}}^{3\rho}J_{R}^{m}\rho\right)\nonumber \\
+ & \sum_{m,m\text{\textasciiacute}=1}^{2}\overline{q_{L}^{m}}\left(h_{D_{R}^{3}}^{m\chi}D_{R}^{3}\chi+h_{J_{R}^{3}}^{m\rho}J_{R}^{3}\rho+h_{D_{R}^{m\text{\textasciiacute}}}^{m\chi}D_{R}^{m\text{\textasciiacute}}\chi+h_{J_{R}^{m\text{\textasciiacute}}}^{m\eta}J_{R}^{m\text{\textasciiacute}}\eta\right)+h.c.
\end{align}

Similarly, for the lepton sector we have

\begin{eqnarray}
-\mathcal{L}_{Y}^{l} & = & \sum_{n,n'=1}^{3}\overline{l_{L}^{n}}\left(h_{\nu_{R}^{n'}}^{n\eta}\nu_{R}^{n'}\eta+h_{e_{R}^{n'}}^{n\rho}e_{R}^{n'}\rho+h_{E_{R}^{n'}}^{n'\chi}E_{R}^{n'}\chi\right)\nonumber \\
 & + & \mathcal{L}_{\pm1/\sqrt{3}}^{l}+\mathcal{L}_{-\sqrt{3}}^{l}+h.c.\label{eq:yukawaleptons}
\end{eqnarray}

with the mixing terms

\begin{equation}
\mathcal{L}_{1/\sqrt{3}}^{l}=\sum_{n=1}^{3}\overline{l_{L}^{n}}\left(h_{e_{R}^{n}}^{n\chi}e_{R}^{n}\chi+h_{E_{R}^{n}}^{n\rho}E_{R}^{n}\rho\right),
\end{equation}

\begin{equation}
\mathcal{L}_{-1/\sqrt{3}}^{l}=\sum_{n=1}^{3}\overline{l_{L}^{n}}\left(h_{\nu_{R}^{n}}^{n\chi}\nu_{R}^{n}\chi+h_{E_{R}^{n}}^{n\eta}E_{R}^{n}\eta\right)+\frac{1}{2}\sum_{n,n'=1}^{3}\overline{l_{L}^{in}}\left(l_{L}^{jn\text{\textasciiacute}}\right)^{c}(h_{\rho}\rho_{k})\epsilon^{ijk},
\end{equation}

\begin{equation}
\mathcal{L}_{-\sqrt{3}}^{l}=\frac{1}{2}\sum_{n,n'=1}^{3}\overline{l_{L}^{in}}\left(l_{L}^{jn\text{\textasciiacute}}\right)^{c}(h_{\eta}\eta_{k})\epsilon^{ijk}.
\end{equation}

From the Yukawa Lagrangians in Eqs. (\ref{eq:yuakwaquarks}-\ref{eq:yukawaleptons})
we obtain the relevant couplings of the $\xi_{\chi}$ component of
the scalar triplet $\chi$ with the new quarks $T$ and $J$ (Table
\ref{tab:fermionsloop}).

\begin{figure}[t]
\begin{centering}
\subfloat[]{\begin{centering}
\includegraphics[scale=0.8]{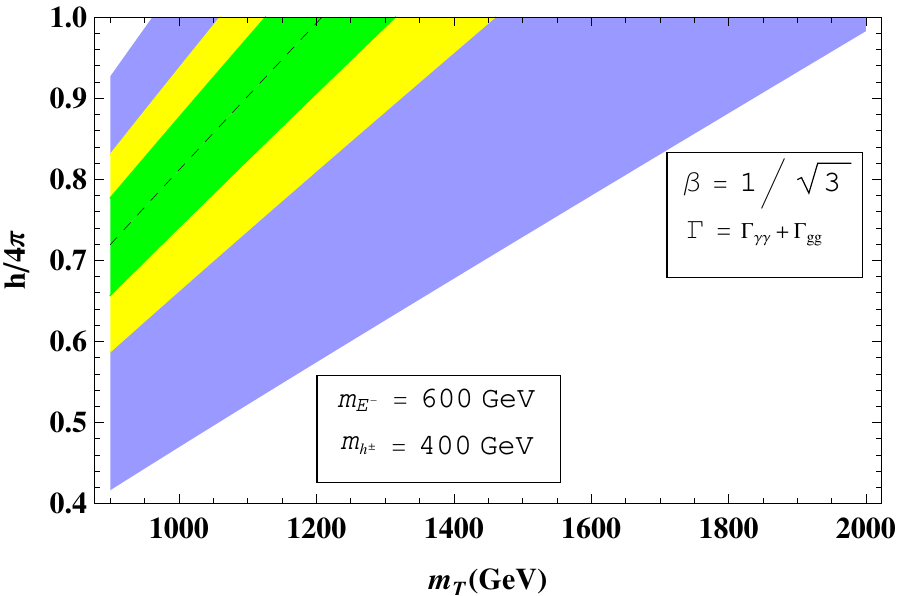}
\par\end{centering}

}$\qquad$\subfloat[]{\begin{centering}
\includegraphics[scale=0.8]{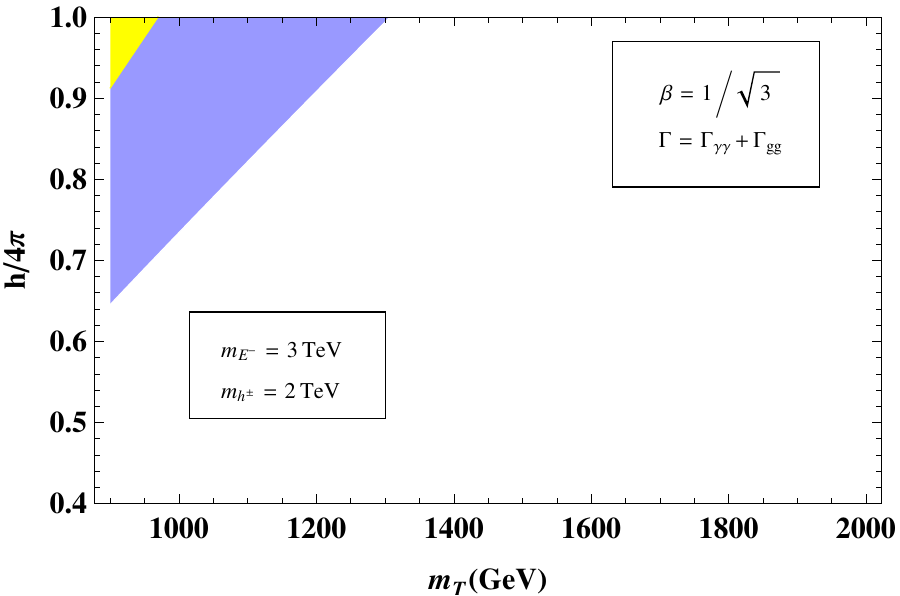}
\par\end{centering}

}
\par\end{centering}

\begin{centering}
\subfloat[]{\begin{centering}
\includegraphics[scale=0.8]{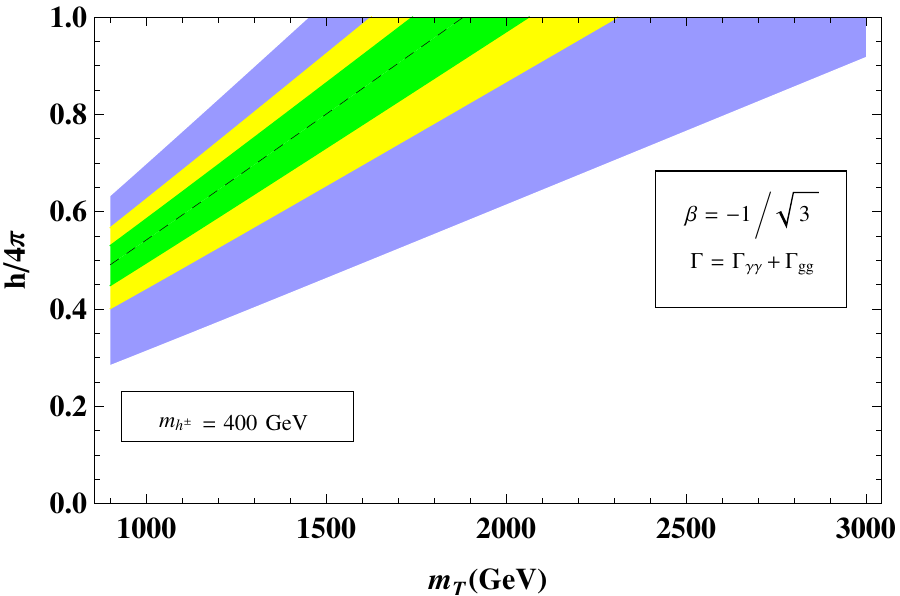}
\par\end{centering}

}$\qquad$\subfloat[]{\begin{centering}
\includegraphics[scale=0.8]{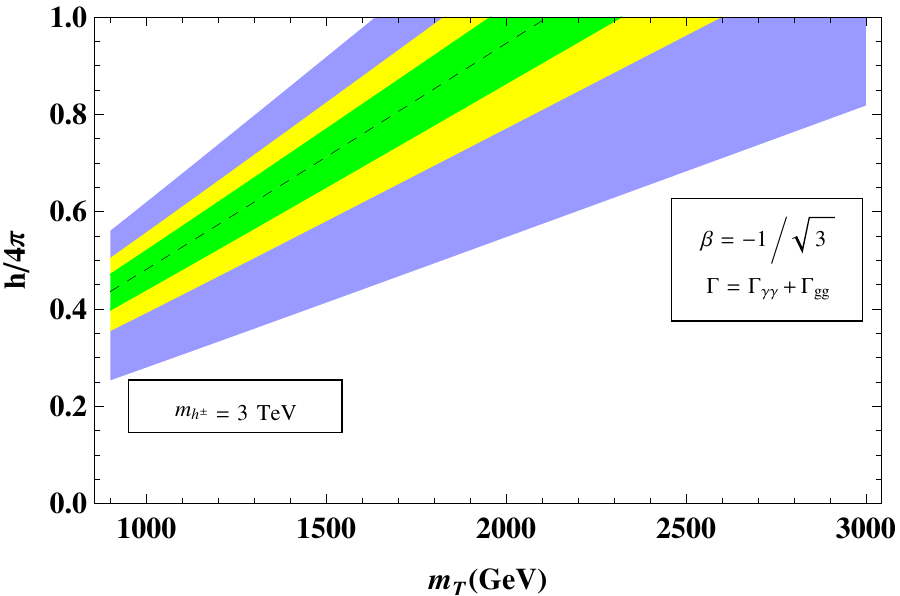}
\par\end{centering}

}
\par\end{centering}

\caption{Contours of the production cross-section $\sigma(pp\to\xi_{\chi}\to\gamma\gamma)$
in femtobarns for $\beta=\pm1/\sqrt{3}$. The dashed line corresponds
to the central value at 6 fb, and the shaded bands corresponds to
regions at 68.3\% (green), 95.5\% (yellow) and 99.7\% (light blue)
C.L. exclusion limits from ATLAS and CMS combined data. \label{fig:1/div3} }
\end{figure}

\begin{figure}[t]
\begin{centering}
\subfloat[]{\begin{centering}
\includegraphics[scale=0.8]{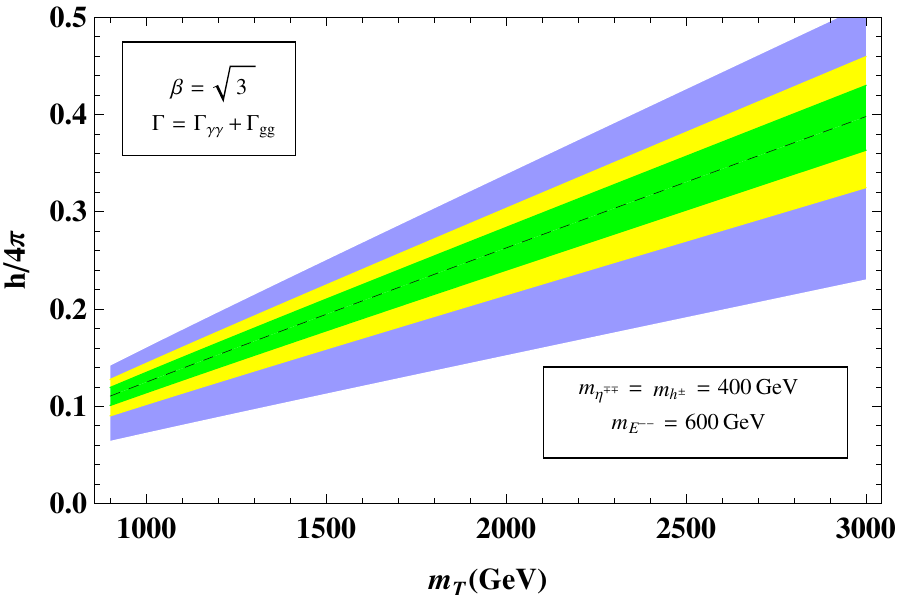}
\par\end{centering}

}$\qquad$\subfloat[]{\begin{centering}
\includegraphics[scale=0.8]{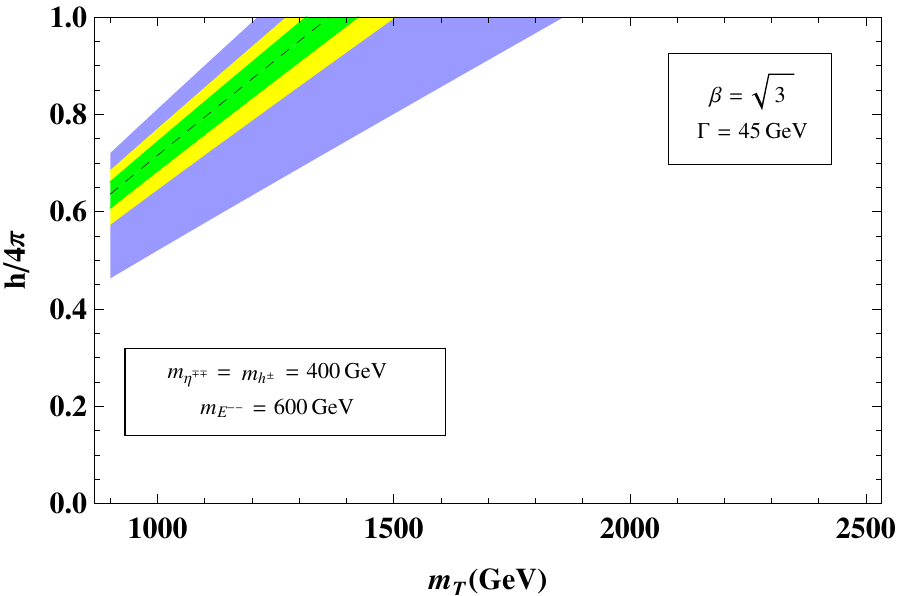}
\par\end{centering}

}
\par\end{centering}

\begin{centering}
\subfloat[]{\begin{centering}
\includegraphics[scale=0.8]{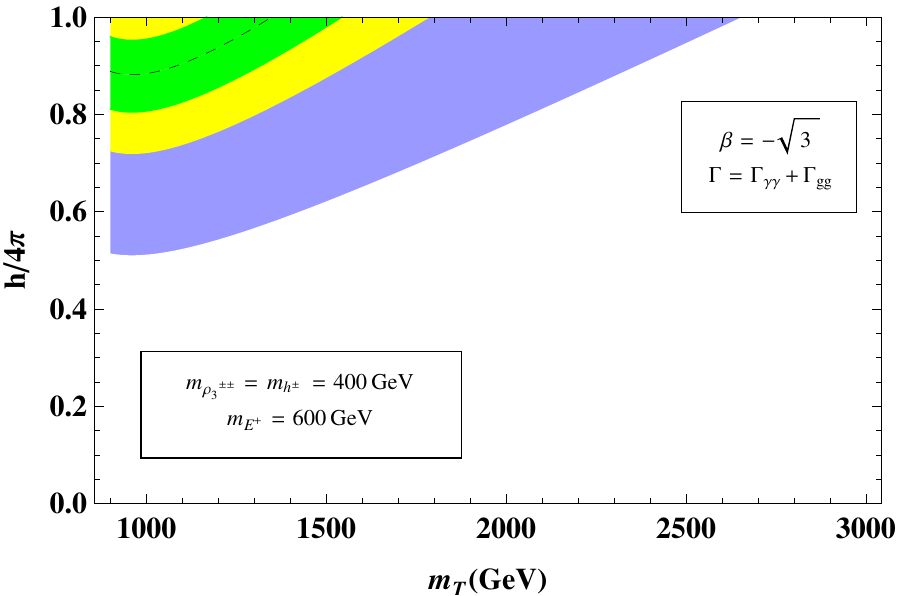}
\par\end{centering}

}$\qquad$\subfloat[]{\begin{centering}
\includegraphics[scale=0.8]{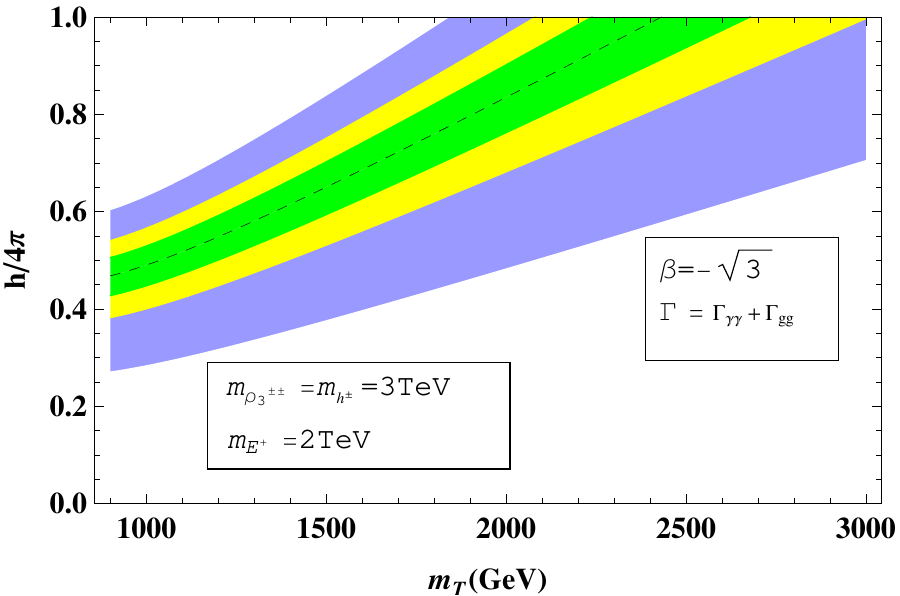}
\par\end{centering}

}
\par\end{centering}

\caption{Contours of the production cross-section $\sigma(pp\to\xi_{\chi}\to\gamma\gamma)$
in femtobarns for $\beta=\pm\sqrt{3}$. The dashed line corresponds
to the central value at 6 fb, and the shaded bands corresponds to
regions at 68.3\% (green), 95.5\% (yellow) and 99.7\% (light blue)
C.L. exclusion limits from ATLAS and CMS combined data. \label{fig:div3}}
\end{figure}

\end{widetext}

\section{Diphoton decay}

We will use two approximations for the decay width: first we assume
one loop contributions only from diphotons and gluons, $\Gamma=\Gamma_{\gamma\gamma}+\Gamma_{gg}$
. Second we take the width given by the experimentally reported value
$\Gamma=45$ GeV from the ATLAS Collaboration. Following \cite{Gunion}
the decay rates of the $\xi_{\chi}$ particle to $\gamma\gamma$ and
$gg$ are 
\begin{eqnarray}
\Gamma(\xi_{\chi}\to\gamma\gamma) & = & \frac{\alpha^{2}h^{2}m_{\xi_{\chi}}^{3}}{512\pi^{3}m_{T}^{2}}\big|\sum_{i}N_{ci}Q_{i}^{2}F_{i}\big|^{2},\\
\Gamma(\xi_{\chi}\to gg) & = & \frac{\alpha_{s}^{2}h^{2}m_{\xi_{\chi}}^{3}}{64\pi^{3}m_{T}^{2}}\big|\sum_{i}F_{i}\big|^{2}
\end{eqnarray}
where $h$ is the Yukawa coupling of the exotic quarks, $m_{T}$ is
the mass of the exotic quarks (assuming the same Yukawa couplings
and masses for simplicity, $m_{T}=m_{J}$), $m_{\xi_{\chi}}$ is the
mass of the scalar candidate, $N_{ci}Q_{i}^{2}$ is the color multiplicity
times the square electric charge and 
\begin{equation}
F_{i}(\tau_{i})=\begin{cases}
2+3\tau_{i}+3\tau_{i}(2-\tau_{i})f(\tau_{i}) & i=1\\
-2\tau_{i}\left[1+(1-\tau_{i})f(\tau_{i})\right] & i=1/2\\
\frac{1}{2}\tau_{i}\left[1-\tau_{i}f(\tau_{i_{i}})\right] & i=0
\end{cases}\label{eq:formfactors}
\end{equation}

\noindent are spin dependent functions for the loop factor. For $\tau_{i}>1$
the function $f(\tau_{i})$ is $f(\tau_{i})=\left[\arcsin\left(\frac{1}{\sqrt{\tau_{i}}}\right)\right]^{2}$
with $\tau_{i}=4m_{i}^{2}/m_{\xi_{\chi}}^{2}$ from where the masses
of the particles into the loop are $m_{i}>375$ GeV . The total cross
section $\sigma(pp\to\xi_{\chi}\to\gamma\gamma)$ in the narrow width
approximation is given by

\begin{equation}
\sigma(pp\to\xi_{\chi}\to\gamma\gamma)=\frac{C_{gg}\Gamma(\xi_{\chi}\to gg)\Gamma(\xi_{\chi}\to\gamma\gamma)}{s\:m_{\xi_{\chi}}\Gamma}
\end{equation}
where $C_{gg}$ is the dimensionless partonic integral computed for
a resonance $m_{\xi_{\chi}}=750$~GeV evaluated at the scale $\mu=m_{\xi_{\chi}}$
and center of mass energy $\sqrt{s}=13\,\mathrm{TeV}$, $C_{gg}=2137$
\cite{Cgg}. 

Here, we have taken $m_{K^{\pm Q_{1}}}=m_{K^{\pm Q_{2}}}\sim3$ TeV
according to ATLAS and CMS Collaborations searches \cite{CMS ATLAS W' mass}.
However, for $m_{K^{\pm Q_{1}}}=m_{K^{\pm Q_{2}}}\sim3$ TeV the associated
form factor $F_{1}$ reaches its asymptotic value, and the cross section
dependence on $m_{K^{\pm Q_{1}}}$ and $m_{K^{\pm Q_{2}}}$ is not
appreciable. So, the production cross section will depend only on
the Yukawa coupling $h$, the mass of the quarks $m_{T}$ and on the
exotic charged lepton masses $m_{E^{-}}$, $m_{E^{--}}$. From the
lower bound reported by the ATLAS Collaboration searches on exotic
heavy charged leptons \cite{lepton masses} we set $m_{E^{-}}=m_{E^{--}}\sim600$
GeV. For the analysis we take the combined results for the cross section
from CMS and ATLAS, $\sigma(pp\to\xi_{\chi}\to\gamma\gamma)=(2-8)\:\mathrm{fb}$
\cite{tools}.

Taking into account all the above conditions, we display in Figs.\ref{fig:1/div3}-\ref{fig:div3}
contour plots of the production cross-section $\sigma(pp\to\xi_{\chi}\to\gamma\gamma)$
as function of the Yukawa coupling normalized as $h/4\pi$ and on
the top-like quark mass $m_{T}$. The lower bound of 900 GeV for $m_{T}$
corresponds to the reported value in recent searches on top- and bottom-like
heavy quarks from ATLAS and CMS Collaborations \cite{quark masses}
and the upper bound of 3 TeV corresponds to the asymptotic value obtained
from the fermionic form factor $F_{1/2}$. We have also taken for
simplicity $m_{h^{\pm}}\equiv m_{\eta_{3}^{\pm}}=m_{\rho_{3}^{\pm}}=400$
GeV and $m_{\eta_{3}^{\pm\pm}}=m_{\rho_{3}^{\pm\pm}}=400$ GeV which
corresponds to the lowest bound from charged Higgs boson searches
reported by ATLAS and CMS \cite{charged higgs} while for the upper
bound we have used $m_{h^{\pm}}=m_{\eta_{3}^{\pm\pm}}=m_{\rho_{3}^{\pm\pm}}=3$
TeV associated to the asymptotic value obtained from the bosonic form
factor $F_{0}$.

In general, from Figs.\ref{fig:1/div3}-\ref{fig:div3} we can observe
that the Yukawa couplings for the smallest masses and positive values
of $\beta$ are larger than for the masses in the asymptotic values.
Conversely, for negative values of $\beta$ the larger the mass parameters,
the smaller the Yukawa couplings. Furthermore, every model exhibits
an allowed region for the diphoton production cross section when $\Gamma=\Gamma_{\gamma\gamma}+\Gamma_{gg}$.
In contrast, for the case $\Gamma=45$ GeV there are allowed regions
only for $\beta=\sqrt{3}$.

Particularly, for $\beta=1/\sqrt{3}(-1/\sqrt{3})$ and mass lower
bounds choices, the model is excluded for Yukawa couplings $h/4\pi<0.4\,(0.3)$
and masses of the exotic quarks $m_{T}>2\,\mathrm{TeV}\,(3\,\mathrm{TeV})$.
On the other hand, for the asymptotic values, the model is excluded
for Yukawa couplings $h/4\pi<0.7\,(0.3)$ and masses of the exotic
quarks $m_{T}>1.3\,\mathrm{TeV}\,(3\,\mathrm{TeV})$. Also, for $\beta=\sqrt{3}(-\sqrt{3})$
and lower bound choices, the model is excluded for Yukawa couplings
$h/4\pi<0.1\,(0.6)$ and masses of the exotic quarks $m_{T}>2.5\,\mathrm{TeV}\,(5\,\mathrm{TeV})$.
For $\beta=\sqrt{3}$ and $\Gamma=45$ GeV, the model is excluded
for Yukawa couplings $h/4\pi<0.5$ and masses of the exotic quarks
$m_{T}>1.8\,\mathrm{TeV}$. Since the exotic quarks and charged Higgs
bosons have electric charge $5/3$ and -2 respectively, the model
for $\beta=\sqrt{3}$ exhibits the smallest Yukawa coupling values
(Fig. \ref{fig:allbetas}).

\subsection{Interference effects}

From Eq.(\ref{eq:formfactors}) we can see a sign difference between
the fermionic and bosonic contributions into the loop for the diphoton
decay. This difference is responsible for interference effects that
can affect considerably the production cross section as shown in Figs
\ref{fig:interference}-\ref{fig:interference4}.

We show in Figs. \ref{fig:interference}, \ref{fig:interference2},
\ref{fig:interference3} and \ref{fig:interference4} the interference
effects from the different contributions for the 331 model for $\beta=1/\sqrt{3},\,-1/\sqrt{3},\,\sqrt{3}$
and $-\sqrt{3}$ respectively. In general, for all $\beta$ values
we can observe the largest interference effects when we only take
into account the vector bosons $K$ (red dotted lines) and the smallest
interference effects arise from the charged Higgs bosons $h^{\pm}$
or $h^{\pm\pm}$ (black dashed lines). Also, we can see that for the
contribution coming only from fermions (green line), we obtain that
the Yukawa couplings take smaller values than when we take into account
all the particles into the loop (blue dot-dashed lines) except for
$\beta=-1/\sqrt{3}$.

Particularly, for model A and $\beta=1/\sqrt{3}$ (Fig \ref{fig:interference})
taking into account only the fermionic contributions (green line)
we obtain allowed regions for $m_{T}$ in agreement with LHC limits.
Also, if we add the charged Higgs boson into the loop (black dashed
line) we obtain allowed regions with $m_{T}>900$ GeV . In contrast,
if we take into account the contribution of the gauge boson $K^{\pm}$,
it produces a strong effect on the production cross section (red dotted
line) excluding the model for the allowed values of $m_{T}>900$ GeV.
However, if we add all the contributions into the loop (blue dot-dashed
line) there appears an allowed region for $m_{T}>900$ GeV.

\begin{widetext}

\begin{figure}[t]
\begin{centering}
\subfloat[]{\begin{centering}
\includegraphics[scale=0.8]{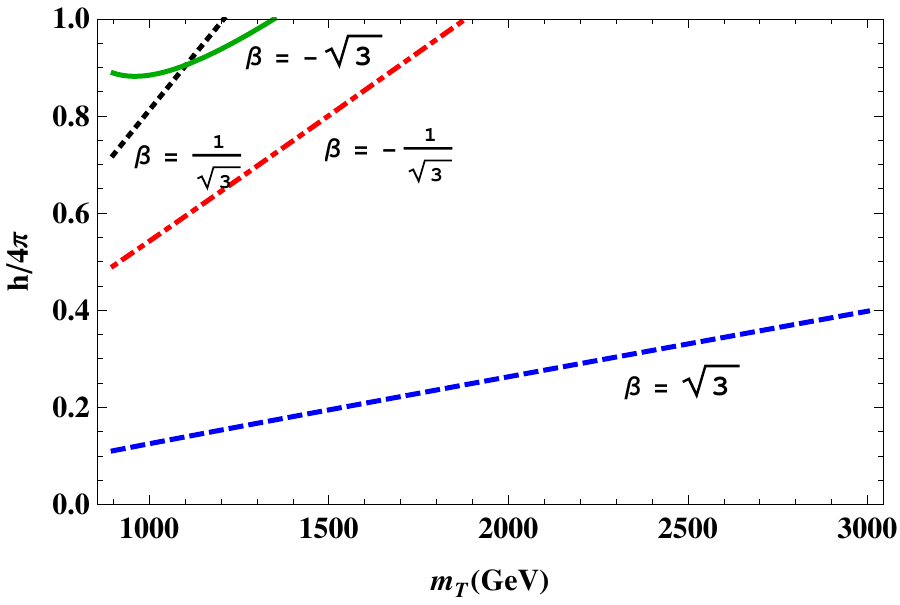}
\par\end{centering}

}$\qquad$\subfloat[]{\begin{centering}
\includegraphics[scale=0.8]{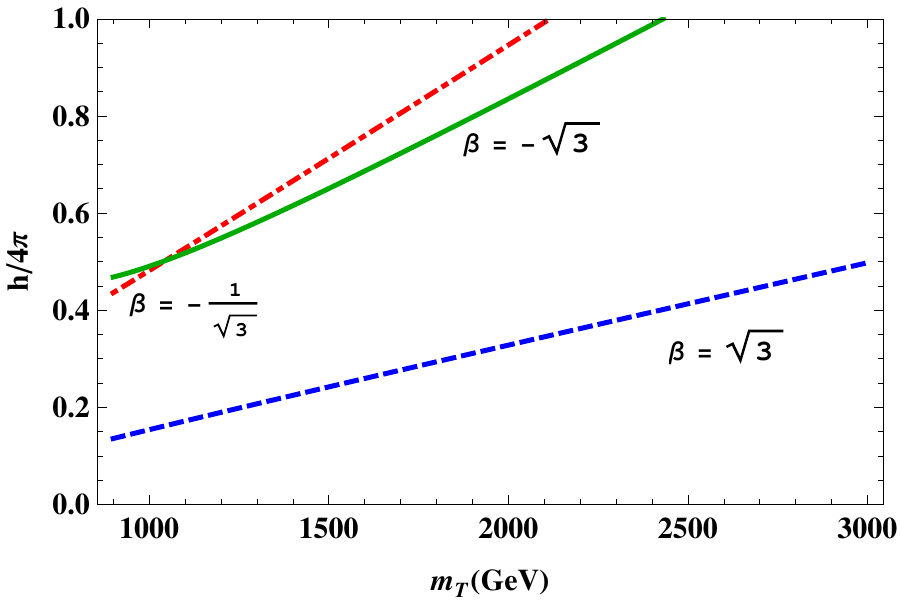}
\par\end{centering}

}
\par\end{centering}

\caption{Contours of the production cross-section $\sigma(pp\to\xi_{\chi}\to\gamma\gamma)$
in femtobarns for the best fit value of 6 fb for (a) mass lower bounds
values reported by LHC and (b) asymptotic mass values coming from
the form factors $F_{i}$ with $i=0,\,1/2,\,1$. \label{fig:allbetas}}
\end{figure}

\begin{figure}[t]
\begin{centering}
\subfloat[]{\begin{centering}
\includegraphics[scale=0.8]{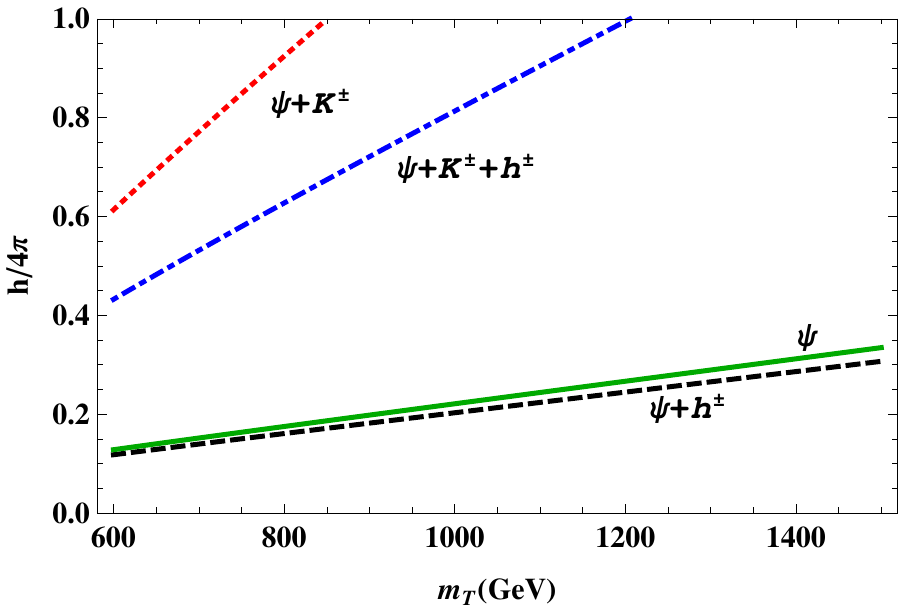}
\par\end{centering}

}$\qquad$\subfloat[]{\begin{centering}
\includegraphics[scale=0.8]{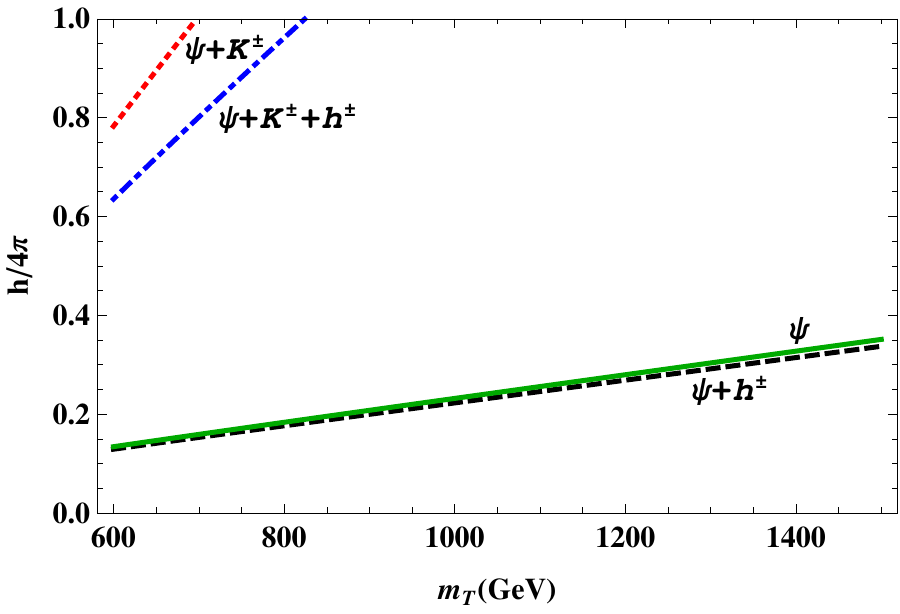}
\par\end{centering}

}
\par\end{centering}

\caption{Contours of the production cross-section $\sigma(pp\to\xi_{\chi}\to\gamma\gamma)$
in femtobarns for the best fit value of 6 fb for (a) mass lower bounds
values reported by LHC and (b) asymptotic mass values coming from
the form factors $F_{i}$ with $i=0,\,1/2,\,1$. Here $\beta=1/\sqrt{3}$
for model A with $\psi=(T^{m},\,J,\,E^{-})$. The green (thick), black
(dashed), blue (dot-dashed) and red (dotted) lines correspond to the
contributions coming from $\psi$, $\psi+h^{\pm}$, $\psi+K^{\pm}+h^{\pm}$
and $\psi+K^{\pm}$ respectively. \label{fig:interference}}
\end{figure}

\begin{figure}[t]
\begin{centering}
\subfloat[]{\begin{centering}
\includegraphics[scale=0.8]{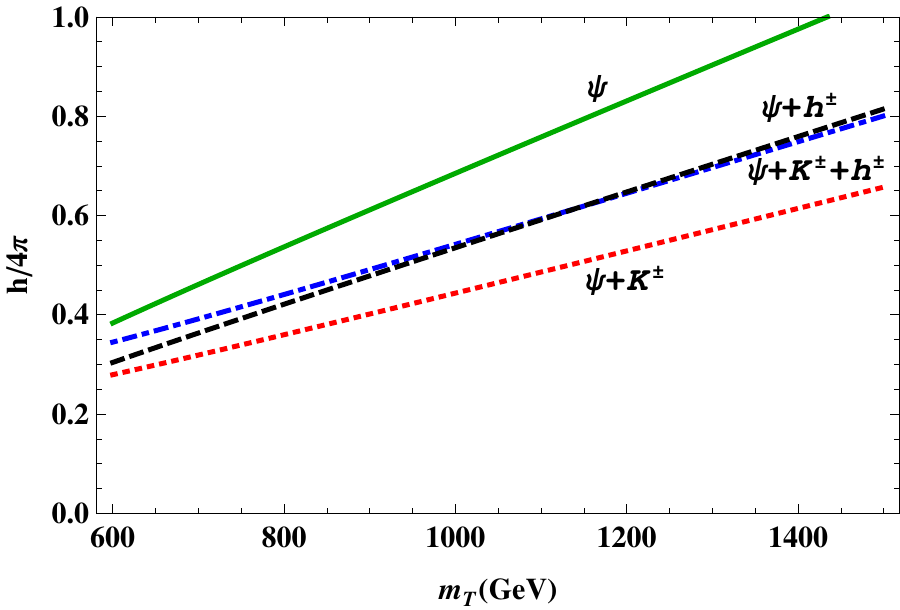}
\par\end{centering}

}$\qquad$\subfloat[]{\begin{centering}
\includegraphics[scale=0.8]{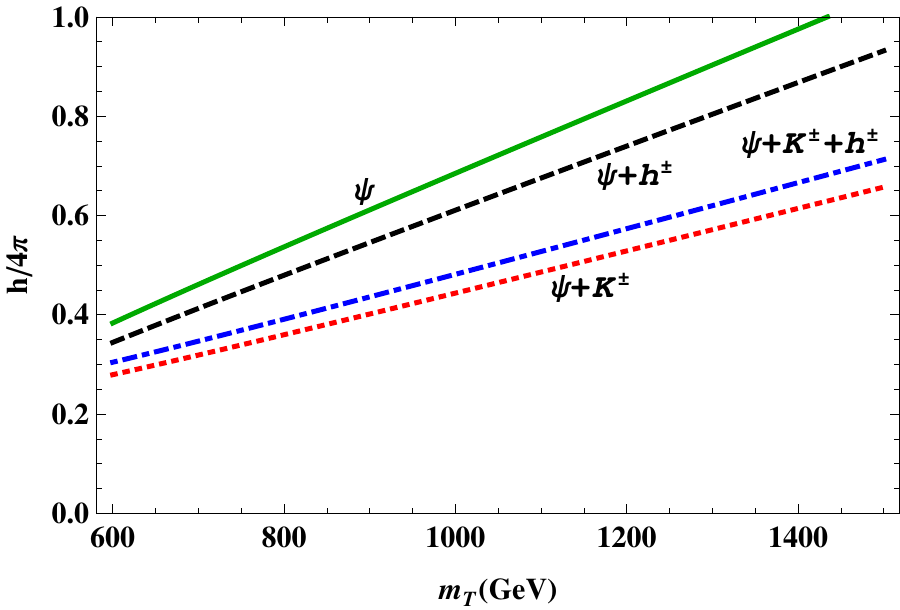}
\par\end{centering}

}
\par\end{centering}

\caption{Contours of the production cross-section $\sigma(pp\to\xi_{\chi}\to\gamma\gamma)$
in femtobarns for the best fit value of 6 fb for (a) mass lower bounds
values reported by LHC and (b) asymptotic mass values coming from
the form factors $F_{i}$ with $i=0,\,1/2,\,1$. Here $\beta=-1/\sqrt{3}$
for model A with $\psi=(T,\,J^{m})$. The green (thick), black (dashed),
blue (dot-dashed) and red (dotted) lines correspond to the contributions
coming from $\psi$, $\psi+h^{\pm}$, $\psi+K^{\pm}+h^{\pm}$ and
$\psi+K^{\pm}$ respectively. \label{fig:interference2}}
\end{figure}

\begin{figure}[t]
\begin{centering}
\subfloat[]{\begin{centering}
\includegraphics[scale=0.8]{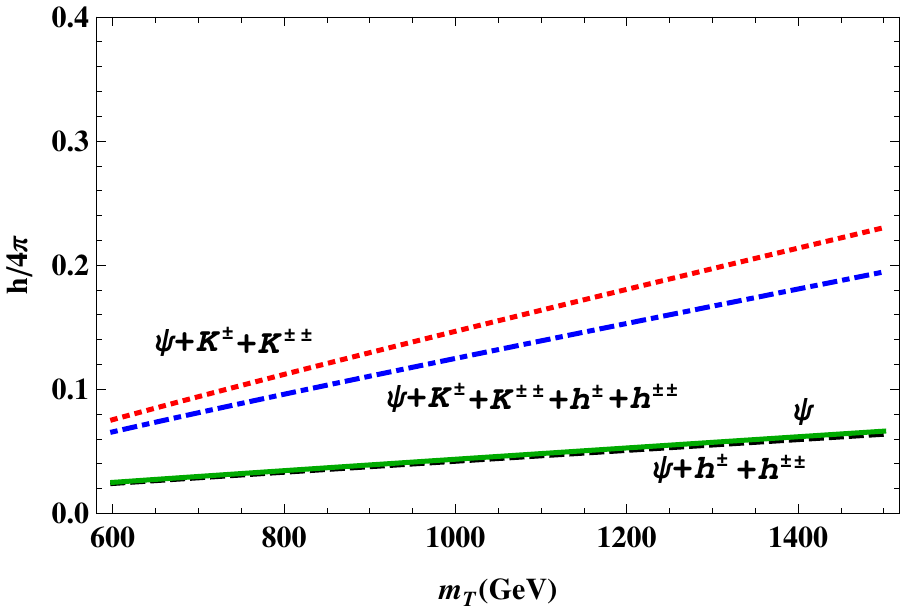}
\par\end{centering}

}$\qquad$\subfloat[]{\begin{centering}
\includegraphics[scale=0.8]{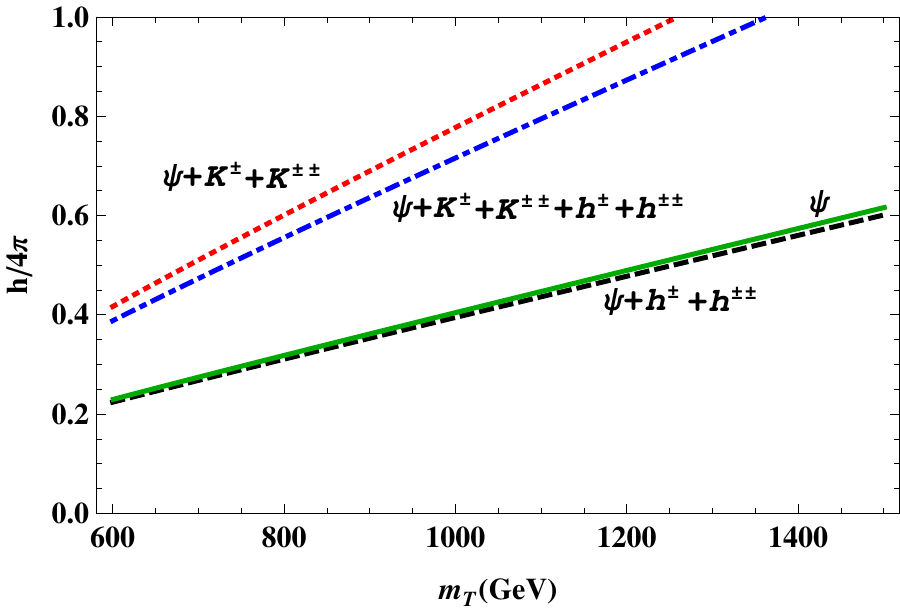}
\par\end{centering}

}
\par\end{centering}

\caption{Contours of the production cross-section $\sigma(pp\to\xi_{\chi}\to\gamma\gamma)$
in femtobarns for the best fit value of 6 fb for (a) mass lower bounds
values reported by LHC and (b) $\Gamma=45$ GeV and same mass lower
bounds. Here $\beta=\sqrt{3}$ for model A with $\psi=(T^{m},\,J,\,E^{--})$.
The green (thick), black (dashed), blue (dot-dashed) and red (dotted)
lines correspond to the contributions coming from $\psi$, $\psi+h^{\pm\pm}$,
$\psi+K^{\pm}+K^{\pm\pm}+h^{\pm}+h^{\pm\pm}$ and $\psi+K^{\pm}+K^{\pm\pm}$
respectively. \label{fig:interference3}}
\end{figure}

\begin{figure}[t]
\begin{centering}
\subfloat[]{\begin{centering}
\includegraphics[scale=0.8]{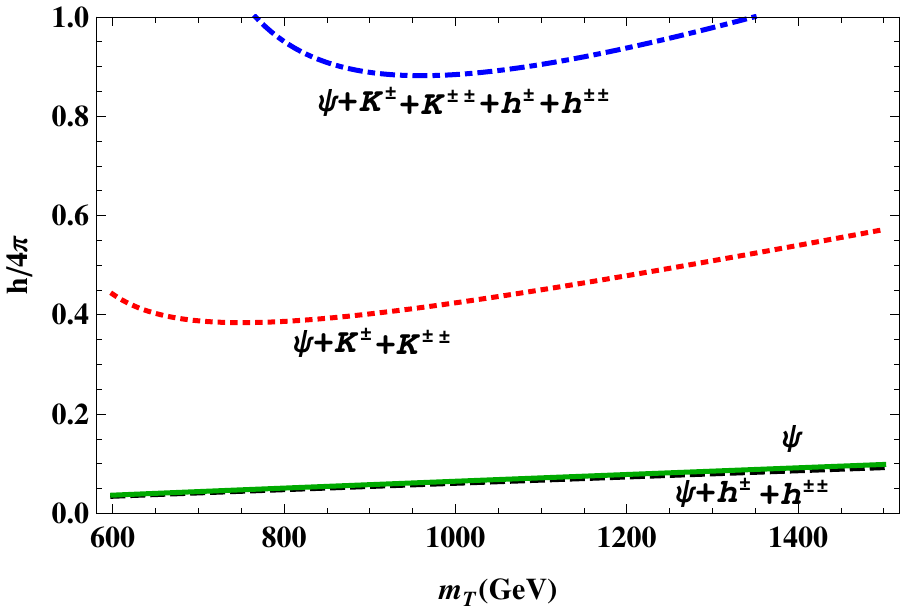}
\par\end{centering}

}$\qquad$\subfloat[]{\begin{centering}
\includegraphics[scale=0.8]{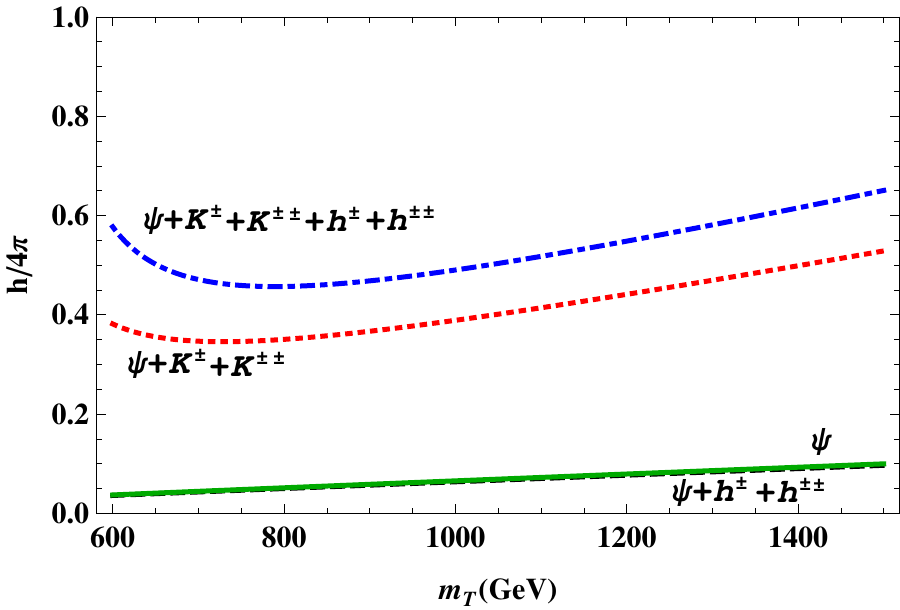}
\par\end{centering}

}
\par\end{centering}

\caption{Contours of the production cross-section $\sigma(pp\to\xi_{\chi}\to\gamma\gamma)$
in femtobarns for the best fit value of 6 fb for (a) mass lower bounds
values reported by LHC and (b) asymptotic mass values coming from
the form factors $F_{i}$ with $i=0,\,1/2,\,1$. Here $\beta=-\sqrt{3}$
for model A with $\psi=(T,\,J^{m},\,E^{+})$. The green (thick), black
(dashed), blue (dot-dashed) and red (dotted) lines correspond to the
contributions coming from $\psi$, $\psi+h^{\pm\pm}$, $\psi+K^{\pm}+K^{\pm\pm}+h^{\pm}+h^{\pm\pm}$
and $\psi+K^{\pm}+K^{\pm\pm}$ respectively. \label{fig:interference4}}
\end{figure}

\section*{Acknowledgments}

This work was supported by El Patrimonio Aut'{o}nomo Fondo Nacional de
Financiamiento para la Ciencia, la Tecnolog\'{i}a y la Innovaci'{o}n
Francisco Jos'{e} de Caldas programme of COLCIENCIAS in Colombia. 

\end{widetext}

\cleardoublepage{}

\end{document}